\documentclass[prb,twocolumn,showpacs,amsmath,amssymb]{revtex4}

\usepackage{graphicx}
\usepackage{latexsym}
\usepackage{amsmath}
\usepackage{amssymb}
\usepackage{amsfonts}
\usepackage{color}

\newcommand{\ds}{\displaystyle}

\begin{document}

\title{Anisotropy and effective dimensionality crossover of the fluctuation conductivity of hybrid superconductor/ferromagnet structures}

\author{S.~V.~Mironov}
\author{A.~S.~Mel'nikov}
\affiliation{Institute for Physics of Microstructures, Russian
Academy of Science, 603950, Nizhniy Novgorod, GSP-105, Russia}

\date{\today}

\begin{abstract}
We study the fluctuation conductivity of a superconducting film, which is placed to perpendicular non-uniform magnetic field with the amplitude $H_0$ induced by the ferromagnet with domain structure. The conductivity tensor is shown to be essentially anisotropic. The magnitude of this anisotropy is governed by the temperature and the typical width of magnetic domains $d$. For $d\ll L_{H_0}=\sqrt{\Phi_0/H_0}$ the difference between diagonal fluctuation conductivity components $\Delta\sigma_\parallel$ along the domain walls and $\Delta\sigma_\perp$ across them has the order of $\left(d/L_{H_0}\right)^4$. In the opposite case for $d\gg L_{H_0}$ the fluctuation conductivity tensor reveals effective dimensionality crossover from standard two-dimensional $\left(T-T_c\right)^{-1}$ behavior well above the critical temperature $T_c$ to the one-dimensional  $\left(T-T_c\right)^{-3/2}$ one close to $T_c$ for $\Delta\sigma_\parallel$ or to the $\left(T-T_c\right)^{-1/2}$ dependence for $\Delta\sigma_\perp$. In the intermediate case $d\approx L_{H_0}$ for a fixed temperature shift from $T_c$ the dependence $\Delta\sigma_\parallel(H_0)$ is shown to have a minimum at $H_0\sim\Phi_0/d^2$ while $\Delta\sigma_\perp(H_0)$ is a monotonically increasing function.
\end{abstract}


\maketitle

\section{Introduction}\label{Intro}

The fluctuation transport in homogeneous superconductors above the critical temperature $T_c$ has been studied for more than half a century (see, e.g., Ref.~\onlinecite{Varlamov_Larkin} for review). The fluctuation correction $\Delta\sigma$ to the Drude conductivity $\sigma_N$ contains three main contributions, which are singular near the critical temperature $T_c$: (i) the positive Aslamasov-Larkin (AL) correction which corresponds to the contribution of non-equilibrium Cooper pairs with finite lifetime to the charge transport\cite{Glover, Aslamazov_Larkin}, (ii) the Maki-Thompson (MT) correction due to single-particle quantum interference at impurities\cite{Maki, Thompson, Patton, Keller} and (iii) the negative correction due to the decrease in the normal electron density of states (DOS)\cite{Di_Castro}. In case of rather strong electron phase-breaking processes the AL correction dominates in fluctuation conductivity since the MT contribution saturates near the critical temperature\cite{Larkin_Ovchinnikov} while the DOS correction is less singular than the AL one.

In the temperature range $Gi\ll \left(T-T_c\right)/T_c\ll 1$ ($Gi$ is the Ginzburg-Levanyuk number\cite{Levanyuk,Ginzburg,Larkin,Larkin_Ovchinnikov}) the influence of fluctuations on electron transport can be described in the frames of the phenomenological Ginzburg-Landau approach. The spatially averaged AL correction $\Delta\sigma^{\alpha\alpha}$  to the diagonal part of the conductivity tensor (along the axis $\alpha$) can be written in the form
\begin{equation}
\label{startexpr} \Delta \sigma^{\alpha\alpha}=
\frac{\pi e^2 \hbar^3}{8m\xi_0^2V}\sum \limits
_{j,l=0}^{\infty}{\frac{{\hat{v}}^{\alpha}_{jl}{\hat{v}}^{\alpha}_{lj}}
{{\varepsilon}_{j}{\varepsilon}_{l} \left({\varepsilon}_{j}+{\varepsilon}_{l}\right)}}.
\end{equation}
Here $V$ is the system volume, $m$ is the
electron mass, $\xi_0$ is the coherence length at zero temperature, indexes $j$ and $l$ include the full set of quantum numbers characterizing the state of non-equillibrium Cooper pair, the set ${\varepsilon}_{j}=\epsilon\hbar^2/(4m\xi_0^2)+E_{j}$ is defined by the reduced temperature $\epsilon=(T-T_{c})/T_{c}$ as well as the set of
eigenvalues $E_{j}$ of the Hamiltonian
\begin{equation}
\label{hamiltonian}
\hat{H}=\frac{\hbar^2}{4m}{\left(-i\nabla-\frac{2\pi}{\Phi_0} \bf{A}\right)}^{2},
\end{equation}
the values ${\hat{v}}^{\alpha}_{jl}$ are the matrix elements of the velocity projection operator
\begin{equation}\label{V_def}
{\hat{v}}^{\alpha}=\frac{\hbar}{2m}
\left(-i{\nabla}^{\alpha}-\frac{2\pi}{\Phi_0} \bf{A}^{\alpha}\right),
\end{equation}
${\bf A}({\bf r})$ is the vector potential of the magnetic field and $\Phi_0=\pi \hbar c/e$ is the flux quantum.

The Aslamazov-Larkin correction is known to have a power-law singularity near the critical temperature. One of the main features is the dependence of the power exponent on the dimensionality $D$ of the superconductor, namely, $\Delta\sigma(T)\propto\left(T-T_c\right)^{D/2-2}$. Another peculiarity is the sensitivity of $\Delta\sigma$ to weak magnetic field, which destroys non-equilibrium Cooper pairs and changes the critical temperature of superconductor. The presence of magnetic field however does not affect the power exponent of $\Delta\sigma(T)$ in the very vicinity of the critical temperature. Thus this exponent is a fundamental value, which reflects the number of degrees of freedom for fluctuating Cooper pairs.

Note also that for spatially homogeneous superconducting systems the energy spectrum of fluctuating Cooper pairs is isotropic in the momentum space. This results in isotropy of the Aslamazov-Larkin correction to the conductivity.

At the same time there are a lot of systems where the superconductivity nucleation is essentially anisotropic, i.e. superconductivity appears not in the whole sample but in spatially localized regions. In particular, in three-dimensional finite superconductors placed into a uniform magnetic field $H$, which is parallel to the samples' edge, the conditions for the superconducting nucleation near the edge are more favorable compared with the bulk. This results in the effect of surface superconductivity.\cite{Saint_James}

A similar type of localized superconductivity appears in superconductors with twinning planes, where even without external magnetic field the local enhancement of the critical temperature near twins takes place (see Ref.~\onlinecite{Khlyustikov_Buzdin} for review). As a result, in certain temperature range the superconductivity exists in the form of two-dimensional nuclei with the width of the order of the coherence length.

\begin{figure}[t!]
\includegraphics[width=0.47\textwidth]{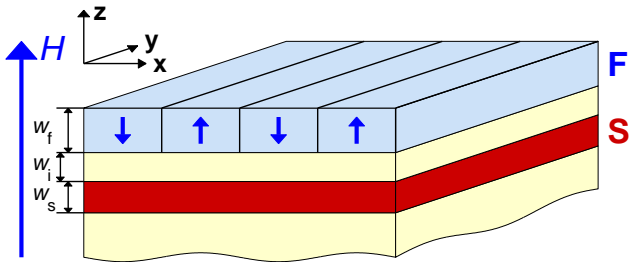}
\caption{(Color online) The planar hybrid structure consisting of a superconducting film and a ferromagnet. The thick ferromagnetic film (F) with domain distribution of magnetization is positioned above a thin superconducting film (S). The system is placed into an external uniform magnetic field ${\bf H}$, applied perpendicular to the surface of the S film.} \label{Fig_SF_Geometry}
\end{figure}

In the past decade a similar phenomenon has been intensively studied in planar hybrid systems which consist of thin superconducting (S) film and a ferromagnet (F) with domain distribution of magnetization (see Fig.~\ref{Fig_SF_Geometry}). Such systems attract growing attention in connection with the possibility to govern transport properties of the S subsystem by manipulating the
domain structure of ferromagnet (see, e.g.,
Ref.~\onlinecite{buzdin,alad} for review). Nucleation of
superconducting state in these systems is strongly affected by
magnitude and spatial configuration of inhomogeneous magnetic
field. It is interesting that for certain parameters of
the system the superconductivity can arise in the form of
separated one-dimensional nuclei which are localized near the
domain walls or inside the domain regions \cite{Buzdin_Melnikov, Aladyshkin_Buzdin}. The experimental evidence of this effect is presented in Refs.~\onlinecite{Yang, Gillijns, Yang_2, Alad_SLM}.

It is interesting that the presence of localized superconducting states can substantially change the transport properties of the superconductors even above $T_c$. In particular, Schmidt and Mikeska\cite{Schmidt_Mikeska} analyzed the fluctuation conductivity of a finite-size superconductor placed into a magnetic field, which is parallel to the samples' edge. They showed that in this system in the very vicinity of the superconducting transition the temperature dependence of the Aslamazov-Larkin correction becomes two-dimensional, i.e.  $\Delta\sigma\propto(T-T_c)^{-1}$. This corresponds to the formation of a narrow two-dimensional channel with enhanced fluctuation conductivity, which is localized near the surface. Later Thompson found that the fluctuation conductivity is anisotropic in the plane of the film due to anisotropy of the effective mass tensor in the spectrum of fluctuating Cooper pairs.\cite{Thompson2} Also he predicted a peculiar dependence of $\Delta\sigma$ on the magnetic field near the tricritical point, where one of the effective mass components changes its sign.
A similar situation is realized in the vicinity of the transition from uniform to the spatially modulated Fulde-Ferrell-Larkin-Ovchinnikov superconducting state.\cite{Konschelle,Konschelle2}. In this case the effective mass in the energy spectrum of the fluctuating Cooper pairs becomes negative and a rich variety of different fluctuation regimes with different critical exponents of the fluctuation conductivity is expected. The effect of the boundary conditions on the anisotropy of $\Delta\sigma$ for a superconductor in the magnetic field, which is parallel to its surface, was analyzed by Imry\cite{Imry}. Also the unusual temperature behavior of $\Delta\sigma$ was predicted for a finite  superconducting film placed into a perpendicular magnetic field.\cite{Zyuzin} In this case near $T_c$ the fluctuation conductivity reveals a one-dimensional behavior due to edge states, i.e. $\Delta\sigma\propto(T-T_c)^{-3/2}$.

A similar anisotropy of fluctuations-dependent quantities appears in the superconductors with twinning planes.\cite{Buzdin_Ivanov} It was shown that if such superconductor is placed into a magnetic field which is perpendicular to the twinning planes then the magnetic susceptibility near the critical temperature has a two-dimensional singularity which is strongly than a three-dimensional one. At the same time, for a longitudinal magnetic field the contribution from the twinning planes to the magnetic susceptibility is negligibly small since the electron motion is appressed to the twins and its experimental observation is very complicated since it is masked by the bulk contribution.

In the present paper we study the fluctuation conductivity of planar hybrid S/F systems in a wide temperature range above $T_c$. We show that in these systems the behavior of the Aslamazov-Larkin correction to conductivity is much more abundant compared to a uniform isolated superconducting film. When the amplitude of the stray magnetic field is zero ($H_0=0$) the dependence of the energy $E$ on the momentum ${\bf k}$ in the plane of the S film has the standard form ($E=\hbar^2 {\bf k}^2/4m$), and the corresponding Aslamazov-Larkin correction to the conductivity $\Delta \sigma$ is isotropic and has standard $\left(T-T_c\right)^{-1}$ singularity at the superconducting transition temperature. In case of finite but small amplitude of the stray field ($H_0d^2\ll\Phi_0$, $d$ is the width of magnetic domains in the ferromagnet) the spectrum stays parabolic for low energies but the effective mass tensor becomes anisotropic, which results in the anisotropy of $\Delta \sigma$ in the plane of superconducting film: the fluctuation conductivity across domain walls exceeds the one along domain walls. With the increasing of $H_0$ the magnitude of $\Delta\sigma$ anisotropy also increases and at $H_0 d^2\approx \Phi_0$ the energy spectrum changes qualitatively: the effective mass corresponding to the momentum $k_y$ along the domain walls changes its sign and two minima of the energy spectrum at non-zero $k_y$ appear. This results in peculiar non-monotonic dependencies of $\Delta\sigma$ components on $H_0 d^2$ at fixed temperature shift from the transition temperature. Finally for $H_0 d^2\gg \Phi_0$ the effective mass corresponding to the momentum across domain walls tends to infinity. The corresponding fluctuation conductivity tensor becomes essentially anisotropic. In particular, the dependence of the component $\Delta\sigma^{yy}$ along domain walls on temperature reveals a crossover from standard two-dimensional $\left(T-T_c\right)^{-1}$ behavior to the one-dimensional $\left(T-T_c\right)^{-3/2}$ one, which corresponds to the formation of quasi-one-dimensional channels with enhanced fluctuations localized
near domain walls. At the same time, the transverse component has the dependence $\Delta\sigma^{xx}\propto\left(T-T_c\right)^{-1/2}$ near the critical temperature. Also we obtain the dependencies of the fluctuation conductivity on the external magnetic field $H$ and analyze possible fluctuation regimes.

\section{Fluctuation conductivity of hybrid S/F structures}\label{Sec_SF}

Let us consider a planar S/F system, which is shown schematically in Fig.~\ref{Fig_SF_Geometry}. A thin superconducting film of the thickness $w_s\ll\xi_0$ and the area $S$ is separated from the ferromagnetic layer with domain distribution of magnetization by an insulating spacer. The thickness of the spacer is assumed to be large enough to neglect the exchange interaction of magnetic moments with electrons in Cooper pairs but, at the same time, rather small so that the magnetic field can penetrate into the superconducting layer without considerable decay. The vector of magnetization $\bf M$ in the ferromagnetic film is assumed to have the $z-$component, which is perpendicular to the plane $xy$ of the superconducting film. Let us consider only the case when the ferromagnet contains magnetic domains with $M_z=\pm M_0$, which are separated by the equidistant set of parallel domain walls. We choose the $y$ axis directed along the domain walls so that $M_z$ depends only on the $x$ coordinate. The width of domains $d$ is assumed to satisfy the condition $\xi_0\ll d\ll \sqrt{S}$. We assume that the domain walls are well pinned and do not take account of changes
in the domain structure with an increase in $H$.

\begin{figure}[t!]
\includegraphics[width=0.45\textwidth]{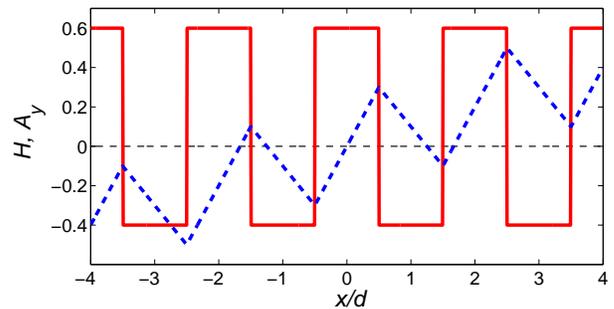}
\caption{(Color online) The spatial profile of the magnetic field $H_z(x)$ (red solid curve) and the corresponding vector potential $A_y(x)$ (blue dashed curve). The parameters are $H_0=0.5H_{c2}^0$ and $H=0.1H_{c2}^0$, where $H_{c2}^{0}= \Phi_0/2\pi{\xi}_{0}^{2}$ .} \label{Fig_Field_Potential}
\end{figure}

In the superconducting film the spatial distribution of the magnetic field $H_z(x)$, induced by the ferromagnet, strongly depends on the thickness $w_f$ of the ferromagnetic layer as well as on the thickness $w_i$ of the insulating spacer between superconducting and ferromagnetic layers. Further we assume that $w_f\gg d$ and $w_i\ll d$. In this case the profile of the stray magnetic field in the superconducting film can be approximated by a meander with the amplitude $H_0=4\pi M_0$:
\begin{equation}\label{Magn_Field}
H_z(x)=H+H_0 {\rm sgn}\left[{\rm cos}\left(\pi x/d\right)\right].
\end{equation}
We choose the corresponding vector potential in the form $A_y(x)=Hx+\tilde{A}_y(x)$, where for any integer $n$
\begin{equation}\label{Vector_Potential}
\frac{\tilde{A}_y(x)}{H_0d}=\left\{\begin{array}{c}{\ds ~~\frac{1}{2}-\left|\frac{x}{d}-2n-\frac{1}{2}\right| ~{\rm for}~2n<\frac{x}{d}<2n+1,}\\{~}\\{\ds -\frac{1}{2}+\left|\frac{x}{d}-2n+\frac{1}{2}\right| ~{\rm for}~2n-1<\frac{x}{d}<2n.}\end{array}\right.
\end{equation}
The spatial profiles of the magnetic field and the vector potential are shown in Fig.~\ref{Fig_Field_Potential}.

\begin{figure}[t!]
\includegraphics[width=0.45\textwidth]{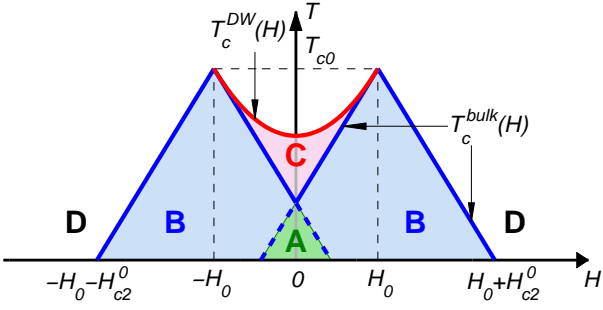}
\caption{(Color online) The phase diagram of a hybrid system consisting of superconducting film and a ferromagnet with domain distribution of magnetization. In the green region (A) for $T<T_c^{CS}(H)$  the superconductivity exists in the whole film. In the blue region (B) for  $T_c^{CS}(H)<T<T_c^{bulk}(H)$ the superconductivity exists only in the domains where the external field and the stray field are contrary directed. In the red region (C) for $\left|H\right|<H_0$ and $T_c^{bulk}(H)<T<T_c^{DW}(H)$ the superconductivity exists in the form of narrow channels which are localized near domain walls. The white area (D) corresponds to the normal state of the film.} \label{Fig_SF_Diagram}
\end{figure}

The phase diagram of the hybrid system under consideration is shown in Fig.~\ref{Fig_SF_Diagram}. For a fixed $H_0$ two different regimes of bulk superconductivity are realized: for $T<T_c^{CS}(H)=T_{c0}\left(1-\left|H_0+\left|H\right|\right|/H_{c2}^0\right)$ (green area (a) in Fig.~\ref{Fig_SF_Diagram}), where $T_{c0}$ is the critical temperature of the isolated superconducting film, the whole sample is superconducting while for $T_c^{CS}(H)<T<T_c^{bulk}(H)= T_{c0}\left(1-\left|H_0-\left|H\right|\right|/H_{c2}^0\right)$ (blue area (b) in Fig.~\ref{Fig_SF_Diagram}) the superconductivity exists only in the regions where the stray field and the external filed compensate each other. An important point is that for $\left|H\right|<H_0$ the superconductivity can exist above $T_c^{bulk}(H)$ in the form of quasi-one-dimensional nuclei localized near the domain walls. This type of localized superconductivity is often called domain-wall superconductivity (red region (c) in Fig.~\ref{Fig_SF_Diagram}). The dependence of the critical temperature $T_c^{DW}$ of domain-wall superconductivity on $H$ is shown schematically by the red curve in Fig.~\ref{Fig_SF_Diagram}.

To calculate the diagonal components of the fluctuation conductivity tensor we use Eq.~(\ref{startexpr}). We will be interested only in spatially averaged correction $\left<\Delta \sigma^{\alpha\alpha}\right>$, which can be obtained by integrating the local correction $\Delta \sigma^{\alpha\alpha}(x)$ over the magnetic domain width:
\begin{equation}
\label{averdef} \left<\Delta \sigma^{\alpha\alpha}\right>=\frac{1}{2d} \int
\limits_{-d}^{d} \Delta \sigma^{\alpha\alpha}\left(x\right)dx.
\end{equation}
Exactly this value determines the drop in the resistance of the sample in transport measurements.

The states of Cooper pairs in the superconducting film are defined by the Schr\"{o}dinger equation
\begin{equation}\label{2D_Schrodinger}
-\partial_x^2\psi+\left(-i\partial_y-\frac{2\pi}{\Phi_0}A_y(x)\right)^2\psi =\frac{4m}{\hbar^2}E\psi.
\end{equation}
At domain walls (at $x=nd$) one should demand the continuity of the order parameter $\psi$ and its derivative $\partial_x\psi$. An arbitrary wave function, which satisfies Eq.~(\ref{2D_Schrodinger}) can be written in the form
\begin{equation}\label{2DFilm_WaveFunction}
\psi(x,y)=\chi_{n,k_y}(x)e^{ik_yy},
\end{equation}
where $k_y$ is the momentum along the $y$ axis and $n$ is a band index. The function $\chi_{n,k_y}(x)$ satisfies the equation
\begin{equation}\label{Film_ChiEquation}
\left[-\partial^2_x +\left(k_y-\frac{2\pi}{\Phi_0}A_y(x)\right)^2\right] \chi_{n,k_y}=\frac{4m}{\hbar^2}E\chi_{n,k_y}.
\end{equation}

\begin{figure*}[hbt!]
\includegraphics[width=0.4\textwidth]{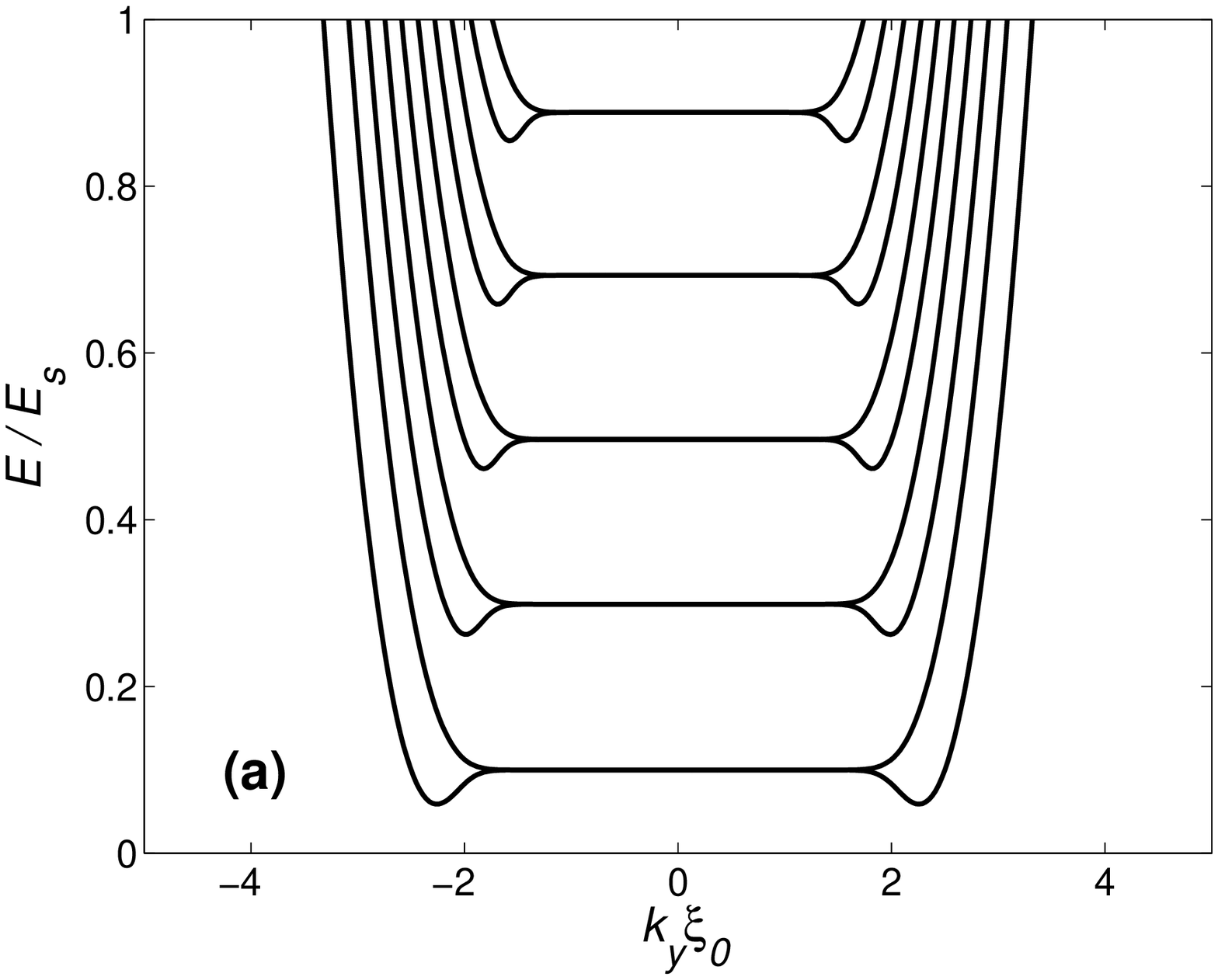}
\includegraphics[width=0.4\textwidth]{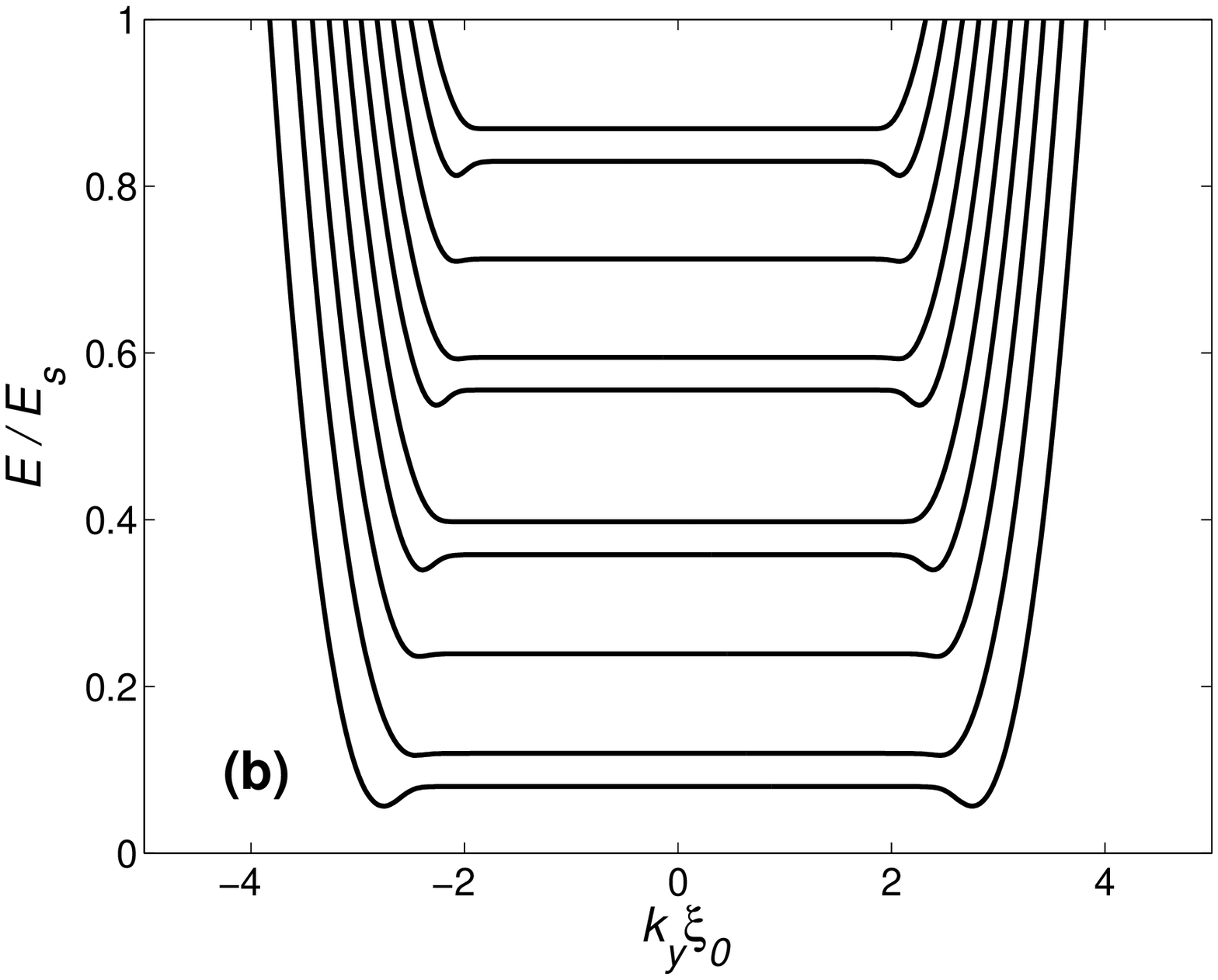}
\caption{The energy spectrum $E(n,k_y)$ of the fluctuating Cooper pairs in the hybrid structure consisting of a superconducting film and a ferromagnet with domain distribution of magnetization, which is placed into an external uniform magnetic field $H$. We take the amplitude of the stray magnetic field $H_0=0.1H_{c2}^0$, the external field $H=0$ (a) and $H=0.02H_{c2}^0$ (b). The width of domains is $d=50\xi_0$ and $E_s=\hbar^2/4m\xi_0^2$.} \label{Fig_2D_spectrum}
\end{figure*}

The form of the spectrum $E$ and the corresponding behavior of the fluctuation conductivity strongly depend on the values of $H$ and $H_0$. In the absence of the ferromagnet (for $H_0=0$) one obtains a standard two-dimensional temperature dependence of $\Delta\sigma$
\begin{equation}
\label{homogfield} \Delta
\sigma=\frac{e^{2}}{2\hbar b}\frac{1}{\epsilon}F\left(\frac{\epsilon}{2h}\right).
\end{equation}
Here $h=H/H_{c2}^{0}$, $H_{c2}^{0}= \Phi_0 /2\pi{\xi}_{0}^{2}$ is the second
critical field at zero temperature, the function $F(x)$ is defined as follows
\begin{equation}
\label{function}
F(x)=x^{2}\left[\Psi\left(\frac{1}{2}+x\right)-\Psi\left(x\right)-\frac{1}{2x}\right],
\end{equation}
and $\Psi$ is the Digamma function. At the same time for $H_0\neq 0$ the situation changes dramatically. Further we analyze three limiting cases corresponding to weak ($H_0 d^2\ll\Phi_0$), strong ($H_0 d^2\gg\Phi_0$) and intermediate ($H_0 d^2\approx\Phi_0$) values of the stray magnetic field.

\subsection{Weak magnetic field}\label{Sec_Weak_Field}

In this section the calculations are base on the following assumptions: (i) the amplitude value $H_0$ of the stray magnetic field is rather small so that the magnetic length $L_{H_0}=\sqrt{\Phi_0/\left|H_0\right|}$ exceeds the width of magnetic domains, i.e.  $L_{H_0}\gg d$; (ii) the temperature is close to the critical one so that $\xi_H(T)\gg d$, where  $\xi_H(T)=\xi_0T_{c0}/\sqrt{T-T_c^{bulk}(0)}$.

We will start from the simplest case when the external magnetic field $H=0$. The magnetic field (\ref{Magn_Field}) in the superconducting film can be expanded into the Fourier series
\begin{equation}\label{Fourier_Def}
H_z(x)=\sum\limits_{n\neq 0}H_n e^{in\pi x/d},
\end{equation}
where $H_n=0$ for $n=2l$ ($l$ is an integer number) and $H_n=2H_0(-1)^{|l|}/\pi (2l+1)$ for $n=2l+1$. For a small $H_0$ one can calculate the spectrum $E$ using the nearly free electron approximation. Indeed if $H_0=0$ the spectrum has a standard parabolic form $E=\hbar^2\left(k_x^2+k_y^2\right)/4m$. The small periodic magnetic field results in small corrections to this spectrum which can be treated within the second order perturbation theory. The detailed discussion of the nearly free electron approximation for small periodic magnetic fields with an arbitrary configuration can be found in Ref.~\onlinecite{Mironov}. The resulting expression for the spectrum in a meander-like magnetic field (\ref{Magn_Field}) has the form
\begin{equation}\label{Weak_Field_Spectrum_Result}
E(k_x,k_y)=\frac{\hbar^2k_x^2}{4m}+ \frac{\hbar^2k_y^2}{4m^{*}_y} +\frac{2e^2d^2}{\pi^2mc^2}\sum_{n=1}^{\infty} \frac{\left|H_n\right|^2}{n^2},
\end{equation}
where
\begin{equation}\label{Weak_Field_Effective_Mass}
m^{*-1}_y=m^{-1}\left(1-\frac{32d^4}{\pi^2\Phi_0^2} \sum_{n=1}^{\infty} \frac{\left|H_n\right|^2}{n^4}\right)
\end{equation}
is the $y-$component of the effective mass tensor. For the specific meander-like form of the stray field the expression (\ref{Weak_Field_Effective_Mass}) for the effective mass transforms into
\begin{equation}\label{Weak_Field_Effective_Mass_result}
m^{*}_y\approx m\left(1+\frac{2\pi^2H_0^2d^4}{15\Phi_0^2}\right).
\end{equation}
The last term in the expression (\ref{Weak_Field_Spectrum_Result}) leads to the shift in the critical temperature of the superconducting film. The resulting critical temperature $T_c(H_0)$ reads as
\begin{equation}\label{Weak_Field_Tc}
T_c(H_0)=T_{c0}\left(1-\frac{\pi^2H_0^2d^2\xi_0^2}{3\Phi_0^2}\right).
\end{equation}

To calculate the fluctuation correction to the conductivity one should substitute the spectrum (\ref{Weak_Field_Spectrum_Result}) into Eq.~(\ref{startexpr}). It is convenient to perform the corresponding calculations for a general case of the spectrum, which has the form $E=E_{min}+\hbar^2 k_x^2/4m^{*}_x+\hbar^2 k_y^2/4m^{*}_y$. Then the $\alpha$-projection of the Cooper pair velocity reads $v^\alpha(k_\alpha)=\hbar^{-1}\partial E/\partial k_\alpha=\hbar k_\alpha/2m^{*}_\alpha$. Performing the summation over quantum indexes in Eq.~(\ref{startexpr}) one obtains that the most singular part of the fluctuation conductivity has the form
\begin{equation}\label{Effective_Mass_Weak_Field_Result}
\left<\Delta\sigma^{\alpha\alpha}\right>= \frac{\sqrt{m^{*}_x m^{*}_y}}{m^{*}_\alpha} \frac{e^2}{16\hbar w_s\epsilon_H},
\end{equation}
where $m^{*}_x=m$ and $m^{*}_y$ is defined by the expression (\ref{Weak_Field_Effective_Mass_result}). From the Eq.~(\ref{Effective_Mass_Weak_Field_Result}) one can see that in the presence of a weak stray magnetic field the Aslamazov-Larkin correction to the conductivity becomes sligtly anisotropic due to anisotropy of the energy spectrum (\ref{Weak_Field_Spectrum_Result}) in the momentum space. The magnitude of this anisotropy is governed by the stray field value (see Eq.~(\ref{Weak_Field_Effective_Mass_result})):
\begin{equation}\label{Weak_Field_Anisotropy}
\frac{\left<\Delta\sigma^{yy}\right>}{\left<\Delta\sigma^{xx}\right>} =\frac{m^{*}_x}{m^{*}_y}\approx 1-\frac{2\pi^2}{15}\left(\frac{d}{L_{H_0}}\right)^4.
\end{equation}

Now let us turn to the case when the external magnetic field $H\neq 0$. We will assume the magnetic field to be rather small so that $L_H\gg d$ (here $L_{H}=\sqrt{\Phi_0/\left|H\right|}$). As previously we assume that $L_{H_0}\gg d$ to use the nearly free electron approximation. The analysis of the spectrum $E$ in case of non-zero external field $H$ and isotropic effective mass tensor was described in Ref.~\onlinecite{Mironov}. The generalization for the case of anisotropic effective mass tensor leads to the spectrum
\begin{equation}\label{Weak_Field_Spectrum_H_Result}
E_l=\frac{e\hbar \left|H\right|}{2c\sqrt{m^{*}_xm^{*}_y}}\left(2l+1\right) +\frac{2e^2d^2}{\pi^2 m c^2}\sum_{n=1}^{\infty} \frac{\left|H_n\right|^2}{n^2},
\end{equation}
where $l$ indicates the number of Landau level. The matrix elements of the velocity projection operator $\hat{v}^\alpha$ has the form
\begin{equation}\label{Weak_Field_Velocity}
\left|\hat{v}^\alpha_{nl}\right|^2=\frac{e\hbar\left|H\right|}{4c m^{*}_\alpha\sqrt{m^{*}_xm^{*}_y}} \left(n\delta_{n,l+1}+l\delta_{l,n+1}\right).
\end{equation}
The expression for the fluctuation conductivity $\left<\Delta\sigma^{\alpha\alpha}\right>$ can be obtained from Eq.~(\ref{homogfield}) by performing the following transformations: (i) the parameter $\epsilon$ should be replaced by $\epsilon_H\sqrt{m^{*}_xm^{*}_y}/m$ and (ii) the whole expression should be multiplied by the factor $\left(m^{*}_xm^{*}_y/mm^{*}_\alpha\right)$. The resulting expression reads:
\begin{equation}\label{Weak_Field_H_General_Result}
\left<\Delta\sigma^{\alpha\alpha}\right>= \frac{\sqrt{m^{*}_xm^{*}_y}}{m^{*}_\alpha}  \frac{e^{2}}{2\hbar w_s} \frac{1}{\epsilon_H}F\left(\frac{\epsilon_H}{2|\tilde{h}|}\right),
\end{equation}
where
\begin{equation}\label{Weak_Field_H_Tilde}
\tilde{h}=\frac{m}{\sqrt{m^{*}_xm^{*}_y}}\frac{H}{H_{c2}^0}
\end{equation}
and the function $F(x)$ is defined by Eq.~(\ref{function}).

Note that for $H\neq 0$ the superconducting transition occurs at temperature \begin{equation}\label{Weak_Field_H_Tc}
T_c^H=T_c(H_0)-T_{c0}|\tilde{h}|.
\end{equation}
Then in the vicinity of $T_c^H$ (when $\epsilon_H+|\tilde{h}|\ll |\tilde{h}|$) the expression (\ref{Weak_Field_H_General_Result}) takes the form
\begin{equation}
\label{Weak_Field_H_Result_Near_Tc}
\left<\Delta\sigma^{\alpha\alpha}\right>= \frac{\sqrt{m^{*}_xm^{*}_y}}{m^{*}_\alpha}  \frac{e^{2}}{4\hbar w_s} \frac{1}{\epsilon_H+|\tilde{h}|}.
\end{equation}

Thus the effect of a weak spatially periodic magnetic field results mainly in the shift of the critical temperature as well as in the anisotropy of the fluctuation conductivity due to anisotropy of the effective mass tensor.

\subsection{Strong magnetic field}\label{Sec_Strong_Field}

In this section we analyze the opposite case when $H_0$ is rather large so that $L_{H_0}\ll d$. In this case the behavior of the fluctuation conductivity strongly depends on the ratio between $H$ and $H_0$. Here we consider all possible regimes.

In the absence of external magnetic field, i.e. $H=0$, the order parameter wave functions $\chi_{n,k_x,k_y}(x)$ satisfy the Bloch theorem. The corresponding energy spectrum slightly depends on the momentum across domain walls $k_x$ and for $k_x=0$ it is shown in Fig.~\ref{Fig_2D_spectrum}(a). For $\left|k_y\right|\ll d/L_{H_0}^2$ the center of the corresponding cyclotron quasiclassical trajectory lays deep inside the magnetic domain region, and the energy spectrum of the Cooper pair has the form of Landau levels, which are degenerated over $k_y$:
\begin{equation}\label{SF_LocSpectrum}
E_n=\frac{e\hbar\left|H_0\right|}{mc}\left(n+1/2\right).
\end{equation}
At $k_y=k_0$ and $k_y=-k_0$ (where $k_0=\pi d/L_{H_0}^2-\sqrt{0.59\left|h_0\right|}/\xi_0$, $h_0=H_0/H_{c2}^0$) the spectrum has two minima. The corresponding eigenstates are localized near the domain walls at $x=(2n+1)d$ and $x=2nd$ respectively. Finally for $\left|k_y\right|\gg d/L_{H_0}^2$ the energy goes up with $\left|k_y\right|$ increasing.

The averaged fluctuation correction to the conductivity of the hybrid structure can be divided into two terms
\begin{equation}\label{2D_Av_Corr_Def}
\left<\Delta \sigma^{\alpha\alpha}\right>=\left<\Delta \sigma^{\alpha\alpha}_{bulk}\right> +\left<\Delta \sigma^{\alpha\alpha}_{DW}\right>,
\end{equation}
where the value
\begin{equation}\label{2D_Zero_Field_Result_Bulk}
\left<\Delta \sigma^{\alpha\alpha}_{bulk}\right> =
\frac{e^{2}}{2\hbar w_s\epsilon}F\left(\frac{\epsilon}{2|h_{0}|}\right)
\end{equation}
is the isotropic contribution from the regions under the magnetic domains (see Eq.~(\ref{homogfield})) calculated in the local approximation (i.e. under the assumption that the the fluctuation conductivity at any certain point of the sample is affected only by the local magnetic field). The value $\left<\Delta \sigma^{\alpha\alpha}_{DW}\right>$ describes the difference between the exact fluctuation correction $\left<\Delta \sigma^{\alpha\alpha}\right>$ and the expression (\ref{2D_Zero_Field_Result_Bulk}). It corresponds to the contribution from narrow regions with the width of the order of $L_{H_0}$ near the domain walls. From Eq.~(\ref{startexpr}) one can see that the singular part of $\left<\Delta \sigma^{\alpha\alpha}_{DW}\right>$ comes only from the terms with $j$ or $l$ corresponding to the lowest energy band with $n=0$. This singular part can be written in the following form:
\begin{equation}
\label{edgeeffect}
\begin{array}{c}{\ds
\left<\Delta \sigma^{\alpha\alpha}_{DW}\right>=\frac{e^2 \hbar^3}{8\pi m \xi_0^2 w_s d}\left[\int\limits_0^{\infty}
\sum \limits
_{n=1}^{\infty}{\frac{\left(2-\delta_{n0}\right)\left|{\hat{v}}^\alpha_{0n}\right|^2}
{{\varepsilon}_{0}{\varepsilon}_{n} \left({\varepsilon}_{0}+{\varepsilon}_{n}\right)}}dk_y\right.}\\{\ds\left. ~~~~~~~~~~~~~~~~~~~~~~~~~~~~~ -\frac{2\pi^2 d}{L_{H_0}^2}{\frac{2\left|{\hat{v}}^{\alpha\prime}_{01}\right|^2}
{{\varepsilon}_{0}^{\prime}{\varepsilon}_{1}^{\prime} \left({\varepsilon}_{0}^{\prime}+{\varepsilon}_{1}^{\prime}\right)}}\right],}
\end{array}
\end{equation}
where $\delta$ is the Kronecker delta. The last term in this expression corresponds to the dominating part in $\left<\Delta \sigma^{\alpha\alpha}_{bulk}\right>$ (here we denote the values obtained in the local approximation by primes). Note that the result of integration over $k_y$ in the region $\left|k_y\right|\ll d/L_{H_0}^2$ fully coincides with the last term. Thus the value $\left<\Delta \sigma^{\alpha\alpha}_{DW}\right>$ is determined only by the region $\left|k_y\right|\gtrsim d/L_{H_0}^2$. Further analysis of the general expression (\ref{edgeeffect}) is rather difficult, so we will consider only the most interesting limiting case.

Let us introduce $\epsilon_{DW}=\left(T-T_c^{DW}\right)/T_{c0}=\epsilon+0.59\left|h_0 \right|\ll \left|h_0\right|$. In this case the singular part of $\left<\Delta\sigma^{yy}_{DW}\right>$ is defined primary by small region near the absolute minimum of the energy spectrum, where the value $\varepsilon_n(k_y)\ll \left|h_0\right|$. This means that it is enough to perform integration only over the region where $k_y\approx \pm k_0$ in Eq.~(\ref{edgeeffect}). Then near the minimum of the lowest band one can consider the power expansion of the dependence $E_0(k_y)$ instead of the exact spectrum. For $k_y>0$ this expansion can be written in the form (see Ref.~\onlinecite{Schmidt_Mikeska}):
\begin{equation}\label{2D_Spectrum_Expansion}
E_0(k_y)\approx\frac{\hbar^2}{4m\xi_0^2}\left[0.59h_0 +0.58\xi_0^2\left(k_y-k_0\right)^2\right].
\end{equation}

Further it is convenient to analyze different components of the Aslamazov-Larkin conductivity tensor separately. We start from the analysis of the $\left<\Delta\sigma^{yy}\right>$ component. First we calculate only the term in Eq.~(\ref{edgeeffect}) with $n=0$. The corresponding diagonal matrix element of the velocity operator $\hat{v}^y_{00}$ is nonzero and for $k_y>0$
\begin{equation}\label{FS_MinVelocity}
\hat{v}^y_{00}(k_y)=\frac{1}{\hbar}\frac{\partial E_0}{\partial k_y}=\frac{0.58\hbar}{2m}\left(k_y-k_{0}\right).
\end{equation}
Substituting the expressions (\ref{2D_Spectrum_Expansion}) and (\ref{FS_MinVelocity}) into Eq.~(\ref{edgeeffect}) we obtain the corresponding part of $\left<\Delta \sigma^{yy}_{DW}\right>$ near $T_c^{DW}$:
\begin{equation}\label{2D_Zero_Field_Result_DW_Y}
\left<\Delta \sigma^{yy}_{DW}\right> =
\frac{\ds \pi e^{2}{\xi}_{0}\sqrt{0.58}}{\ds
16\hbar w_s d~\epsilon_{DW}^{3/2}}.
\end{equation}

Note that terms with $n \neq 0$ do not make noticeable contribution to $\left<\Delta\sigma^{yy}_{DW}\right>$ and can be neglected since they are less singular than the expression (\ref{2D_Zero_Field_Result_Bulk}). To show this let us assume that the matrix elements $\hat{v}^y_{n0}(k_y)$ do not depend on $k_y$ (strictly speaking it is not true but the consideration of exact expressions for $\hat{v}^y_{n0}(k_y)$ does not change the main conclusion).  Then from the Eq.~(\ref{edgeeffect}) one can see that the corresponding contribution to $\left<\Delta\sigma^{yy}_{DW}\right>$ is proportional to $\epsilon_{DW}^{-1/2}$ and it is less singular than the expression (\ref{2D_Zero_Field_Result_DW_Y}).

Thus the full averaged fluctuation correction to the conductivity has the form
\begin{equation}\label{2D_Zero_Field_Full_Result_DW}
\left<\Delta \sigma^{yy}_{DW}\right> =\frac{e^{2}}{2\hbar w_s\epsilon}F\left(\frac{\epsilon}{2|h_{0}|}\right)+
\frac{\ds \pi e^{2}{\xi}_{0}\sqrt{0.58}}{\ds
16\hbar w_s d~\left(\epsilon+0.59\left|h_0 \right|\right)^{3/2}}.
\end{equation}

One can see that far from the superconducting transition (when $\epsilon_{DW}\gg \left|h_0\right|$) the contribution from the domain walls is negligibly small and the averaged correction to the conductivity coincides with the general expression (\ref{homogfield}) for the two-dimensional case. In the opposite limit $\epsilon_{DW}\ll \left|h_0\right|$ the situation changes. The contribution from the domain regions to the Aslamazov-Larkin correction is not singular at $T_c^{DW}$ and has the order $\left<\Delta \sigma^{yy}_{bulk}\right>\propto e^2/(\hbar w_s\left|h_0\right|)$ while the edge contribution diverges as $\left<\Delta \sigma^{yy}_{DW}\right>\propto e^2\xi_0/(\hbar w_s d) \epsilon_{DW}^{-3/2}$. It is important that near the critical temperature when $\epsilon_{DW}\ll \left[(\xi_0/d)\left|h_0\right|\right]^{2/3}\propto \xi_0^2\left(dL_H^2\right)^{-2/3}$ the correction $\left<\Delta \sigma^{yy}_{DW}\right>$ becomes dominating. Moreover in this temperature region the total averaged correction to the conductivity has the one-dimensional singularity.

\begin{figure}[t!]
\includegraphics[width=0.4\textwidth]{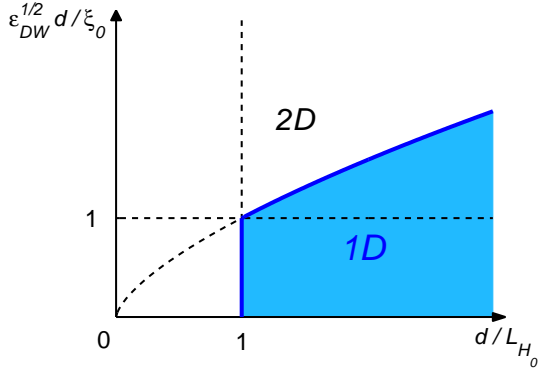}
\caption{(Color online) The diagram of different temperature regimes for fluctuation conductivity along the domain walls in hybrid planar S/F systems in the plane of key parameters. The horizontal axis corresponds to the line of superconducting transition, i.e. to $T=T_c^{DW}(H_0)$. In a white region the singularity of the fluctuation conductivity is two-dimensional and $\Delta\sigma\propto\left(T-T_c\right)^{-1}$ while in dark region it becomes one-dimensional due to the local enhancement of fluctuations near the domain walls and $\Delta\sigma\propto\left(T-T_c\right)^{-3/2}$.} \label{Fig_Global_Diagram}
\end{figure}

The regimes with different behavior of $\left<\Delta\sigma^{yy}\right>$ are shown schematically in Fig.~\ref{Fig_Global_Diagram}. If $d\ll L_{H_0}$ the stray magnetic field is weak and the temperature behavior of $\left<\Delta\sigma^{yy}\right>$ is two-dimensional, i.e. $\left<\Delta\sigma^{yy}\right>\propto \left[T-T_c^{bulk}(H_0)\right]^{-1}$. In the opposite case when $d\gg L_{H_0}$ there is temperature region near the critical temperature of domain-wall superconductivity, where the fluctuations become one-dimensional and $\left<\Delta\sigma^{yy}\right>\propto \left[T-T_c^{DW}(H_0)\right]^{-3/2}$ (the dark region in Fig.~\ref{Fig_Global_Diagram}).

To analyze the $\left<\Delta\sigma^{xx}\right>$ component of the fluctuation conductivity tensor one should notice that the diagonal matrix element $\hat{v}^{x}_{00}$ is exponentially small and can be neglected. The exact calculation of non-diagonal elements is rather cumbersome while for analysis of the dependence of $\left<\Delta\sigma^{xx}\right>$ on temperature it is enough to make a simple estimate. For localized states of fluctuating Cooper pairs $\left|\hat{v}^{x}_{0n}\right|\propto\hbar/mL_{H_0}$ while for delocalized states with high energies these matrix elements are exponentially small. Thus, taking into account the series connection of the regions under the magnetic domains and the regions of domain walls, one can obtain that
\begin{equation}\label{FS_Result_X_Estimate}
\left<\Delta\sigma^{xx}_{DW}\right> \propto\frac{e^2L_{H_0}^2}{\hbar w_s\xi_0d} \frac{1}{\epsilon_{DW}^{1/2}}.
\end{equation}
One can see that for $\epsilon_{DW}\ll \left(\xi_0/d\right)^2$ the contribution from the domain wall regions becomes dominant in the averaged fluctuation conductivity $\left<\Delta\sigma^{xx}_{DW}\right>$.

From the above analysis one can see that the Aslamazov-Larkin conductivity tensor is anisotropic. The striking feature is that the magnitude of this anisotropy strongly depends on temperature. To show this let us compare the expressions (\ref{2D_Zero_Field_Result_Bulk}), (\ref{2D_Zero_Field_Result_DW_Y}) and (\ref{FS_Result_X_Estimate}). For $\epsilon_{DW}\gg \xi_0^2\left(dL_{H_0}^2\right)^{-2/3}$ the fluctuation conductivity tensor is isotropic and its components are determined by the stray magnetic field in the regions under the domains (see Eq.~(\ref{2D_Zero_Field_Result_Bulk})). In the temperature range $\xi_0^2/d^2\ll \epsilon_{DW} \ll \xi_0^2\left(dL_{H_0}^2\right)^{-2/3}$ the fluctuation conductivity tensor becomes anisotropic and $\left<\Delta\sigma^{yy}\right>/\left<\Delta\sigma^{xx}\right> \propto \left(\xi_0^3/L_{H_0}^2d\right)\epsilon_{DW}^{-3/2}$. Finally for $\epsilon_{DW}\ll\xi_0^2/d^2$ the anisotropy is determined by the ratio between the contributions from the domain wall regions and does not depend on $d$: $\left<\Delta\sigma^{yy}\right>/\left<\Delta\sigma^{xx}\right> \propto \left(\xi_0/L_{H_0}\right)^2\epsilon_{DW}^{-1}$. The experimental observation of these temperature crossovers could be a direct illustration of the domain boundary effect on the fluctuation conductivity of hybrid S/F structures.

Now let us turn to the case of intermediate external magnetic field values, i.e. $0<\left|H\right|\ll\left|H_0\right|$. In this case the splitting of the Landau levels occurs due to the difference in the total magnetic field in the neighboring domains (see Fig.~\ref{Fig_2D_spectrum}b). Also the absolute minima of the energy spectrum in this case shift towards the points $k_y=\pm k_0^H$, where $k_0^H>k_0$. Then in the region $\left|k_y\right|\lesssim k_0^H$ the spectrum contains two sets of Landau levels corresponding to the fields $H_0+H$ and $H_0-H$.

The contribution $\left<\Delta \sigma^{\alpha\alpha}_{bulk}\right>$ calculated in the local approximation contains two terms which come from the domains with antiparallel direction of magnetization:
\begin{equation}\label{2D_Result_Bulk}
\left<\Delta \sigma^{\alpha\alpha}_{bulk}\right> =
\frac{e^{2}}{4\hbar w_s\epsilon}\left[F\left(\frac{\epsilon}{2|h+h_{0}|}\right)+
F\left(\frac{\epsilon}{2|h-h_{0}|}\right)\right].
\end{equation}

\begin{figure}[t!]
\includegraphics[width=0.4\textwidth]{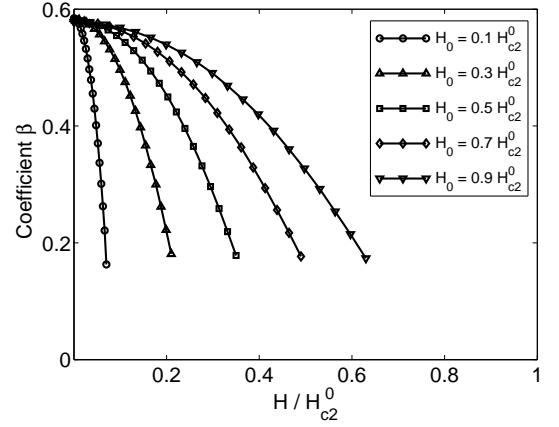}
\caption{The dependence of the coefficient $\beta$, which describes the effective mass of fluctuating Cooper pairs (see Eq.~(\ref{2D_Spectrum_Approx})), on the external magnetic field $H$ for different $H_0$ values.} \label{Fig_Beta}
\end{figure}

In the vicinity of the critical temperature $T_c^{DW}(H)$ of the superconducting transition the contribution $\left<\Delta \sigma^{\alpha\alpha}_{DW}\right>$ is governed by the $k_y$ values which are close to $\pm k_0^{H}$. In this case the lowest band of the spectrum can be approximated by the quadratic function of the form
\begin{equation}\label{2D_Spectrum_Approx}
E_0(k_y)=E_{min}+\frac{\beta\hbar^2}{4m}\left(k_y-k_0^{H}\right)^2.
\end{equation}
The values $E_{min}$ and $\beta$ depend on both $H_0$ and $H$ (they should be obtained from the exact spectrum). The detailed analysis of the dependence $E_{min}(H,H_0)$ is presented in Ref.~\onlinecite{Alad_Moshchalkov}. Here we will focus on the dependencies $\beta(H)$. Performing numerical calculations we have obtained typical dependencies $\beta(H)$ for different $H_0$, which are shown in Fig.~\ref{Fig_Beta}. One can see that at $H=0$ the coefficient $\beta$ does not depend on $H_0$ and is approximately equal to $\beta\approx 0.58$, while the nonzero magnetic field $H$ strongly suppresses the value of $\beta$. It should be mentioned that our analysis is valid for $H<H_0$ only. Then for the diagonal components of the fluctuation conductivity tensor we obtain
\begin{equation}\label{2D_Result_DW}
\left<\Delta \sigma^{yy}_{DW}\right> =
\frac{\ds \pi e^{2}{\xi}_{0}\sqrt{\beta}}{\ds
16\hbar w_sd~\epsilon_{DW}^{3/2}}
\end{equation}
and
\begin{equation}\label{FS_Result_DW_X_Estimate}
\left<\Delta\sigma^{xx}_{DW}\right> \propto\frac{e^2L_{H_0}^2}{\hbar w_s\xi_0d\sqrt{\beta}} \frac{1}{\epsilon_{DW}^{1/2}}.
\end{equation}

The expressions (\ref{2D_Result_DW}) and (\ref{FS_Result_DW_X_Estimate}) are valid for $\epsilon_{DW}\ll\left|h_0-h\right|-\left(4m\xi_0^2/\hbar^2\right)E_{min}(H_0,H)$. Note that dependence of the effective mass of fluctuating Cooper pairs on $H$ leads to an additional field-dependent anisotropy of the Aslamazov-Larkin conductivity tensor since $\left<\Delta \sigma^{yy}_{DW}\right>/\left<\Delta \sigma^{xx}_{DW}\right>\propto\beta(H)$.

Finally for $\left|H\right|>\left|H_0\right|$ the total magnetic field does not change its sign at domain walls and the regime of localized superconductivity can not be realized. At the critical temperature $T_c^{bulk}(H)$ the domains, in which the stray magnetic field and external field have opposite directions, switch to the superconducting state while other domains stay in normal state. Then the total fluctuation correction to the conductivity of the sample is described by the expression (\ref{2D_Result_Bulk}).

\begin{figure}[t!]
\includegraphics[width=0.35\textwidth]{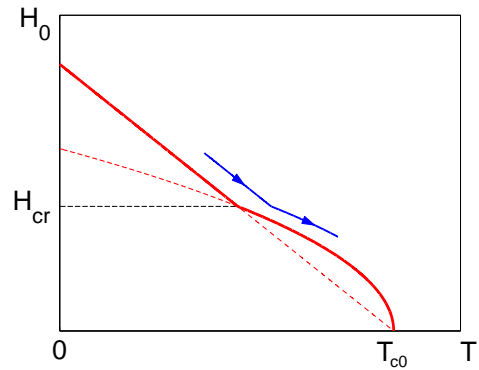}
\caption{(Color online) The phase diagram of a hybrid S/F system in the absence of the external magnetic field. Red solid curve shows the dependence of the critical temperature on the amplitude of the stray magnetic field. At the magnetic field $H_{cr}$ the component of the effective mass along the domain walls changes its sign. Blue solid curve with arrows shows the example of contour with the fixed temperature shift from the transition temperature.} \label{Fig_H0_T_Diagram}
\end{figure}

\subsection{Intermediate magnetic field} \label{Sec_Inter_Field}

In this section we consider the case of intermediate amplitude values of the stray magnetic field, i.e. $H_0\approx\Phi_0/d^2$ ($d\approx L_{H_0}$). At $H_0=H_{cr}=1.02 \Phi_0/d^2$  the energy spectrum of the fluctuating Cooper pairs changes qualitatively: one spectrum minimum at $k_y=0$ for $H_0<H_{cr}$ splits into two minima at finite $k_y$ for $H_0>H_{cr}$.  This splitting results in the break of the transition line at the phase diagram, which is shown in Fig.~\ref{Fig_H0_T_Diagram}. Note that in the absence of external magnetic field the black solid line in Fig.~\ref{Fig_H0_T_Diagram} coincides with the temperature dependence of the upper critical field for a superconducting film in the longitudinal magnetic field.\cite{SJames}

For $H_0\approx H_{cr}$ the energy spectrum of the fluctuating Cooper pairs can be written in the form
\begin{equation}\label{Inter_Spectrum}
E=E_0+\frac{\hbar^2 k_x^2}{4m_x^{*}}+\frac{\hbar^2 k_y^2}{4m_y^{*}}+\eta k_y^4,
\end{equation}
where the minimum $E_0(H_0)$ of the energy spectrum determines the critical temperature of the system. The inverted effective mass component $m_y^{*-1}$ is small and changes its sign with the variation of $H_0$:
\begin{equation}\label{Eff_Mass}
m_y^{*-1}=m_{y0}^{*-1}\left(1-\frac{H_0}{H_{cr}}\right),
\end{equation}
where $m_{y0}^{*-1}$ does not depend on $H_0$. The effective mass component $m^{*}_x$ has the order of the free electron mass $m$ since for $H\gtrsim H_{cr}$ the decaying length of the wave function localized near the domain wall is of the order of $L_{H_0}\approx d$ and the overlapping of the wave functions at neighboring domain walls is essential.

Substituting the spectrum (\ref{Inter_Spectrum}) into Eq.~(\ref{startexpr}) one obtains (the details of calculation can be found in Appendix)
\begin{equation}\label{Inter_X_result}
\left<\Delta\sigma^{xx}\right>= \frac{e^2\sqrt{\xi_0}m^{1/4}}{32 \sqrt{2m_x^{*}}\hbar w_s\eta^{1/4}\epsilon_t^{5/4}}G_x\left(\mu\right),
\end{equation}
\begin{equation}\label{Inter_Y_result}
\left<\Delta\sigma^{yy}\right>=\frac{e^2\sqrt{2m_x^{*}}\eta^{1/4} }{16 \hbar w_s\sqrt{\xi_0}m^{1/4}\epsilon_t^{3/4}}G_y\left(\mu\right)
\end{equation}
where $\epsilon_t=\epsilon+4mE_0/\hbar^2$, the functions $G_x\left(\mu\right)$ and $G_y\left(\mu\right)$ are defined as
\begin{widetext}
\begin{equation}\label{Gx_def}
G_x\left(\mu\right)=\left\{
\begin{array}{l}{\ds \frac{1}{1-\mu}\left[K\left(\frac{1-\mu}{2}\right)- \frac{2\mu}{\left(1+\mu\right)} E\left(\frac{1-\mu}{2}\right)\right] ~~~~~~~~~~~~~~~{\rm for~}-1<\mu<1,}\\{ \ds \frac{1}{ \left(\mu^2-1\right)\sqrt{s}}\left[\mu  E\left(2s\sqrt{\mu^2-1}\right) -s K\left(2s\sqrt{\mu^2-1}\right)\right]~~~{\rm for~}\mu>1,}\end{array}\right.
\end{equation}
\begin{equation}\label{Gy_def}
G_y\left(\mu\right)=\left\{
\begin{array}{l}{\ds \frac{1}{1-\mu}\left[\frac{2\left(3-\mu^2\right)}{ 1+\mu}E\left(\frac{1-\mu}{2}\right)- \left(3-\mu\right)K\left(\frac{1-\mu}{2}\right)\right] ~~~~~~~~~~~~~~{\rm for~}-1<\mu<1,}\\{ \ds \frac{1}{ \left(\mu^2-1\right)\sqrt{s}}\left[2\mu s K\left(2s\sqrt{\mu^2-1}\right) -\left(3-\mu^2\right)E\left(2s\sqrt{\mu^2-1}\right)\right]~~~{\rm for~}\mu>1.}\end{array}\right.
\end{equation}
\end{widetext}
In the expressions (\ref{Inter_X_result})-(\ref{Gy_def}) we denote $\mu=\sqrt{m}\xi_0/ \left(4m_y^{*}\sqrt{\eta\epsilon_t}\right)$, $s=\mu- \sqrt{\mu^2-1}$, $E\left(\nu\right)$ and $K\left(\nu\right)$ are complete elliptic integrals of the first and the second kind respectively ($\left|\nu\right|<1$):
\begin{equation}\label{E_K_Def}
E\left(\nu\right)=\int\limits_0^{\pi/2} \sqrt{1-\nu\sin^2\varphi}d\varphi,K\left(\nu\right)=\int\limits_0^{\pi/2} \frac{d\varphi}{\sqrt{1-\nu\sin^2\varphi}}.
\end{equation}

\begin{figure}[hbt!]
\includegraphics[width=0.23\textwidth]{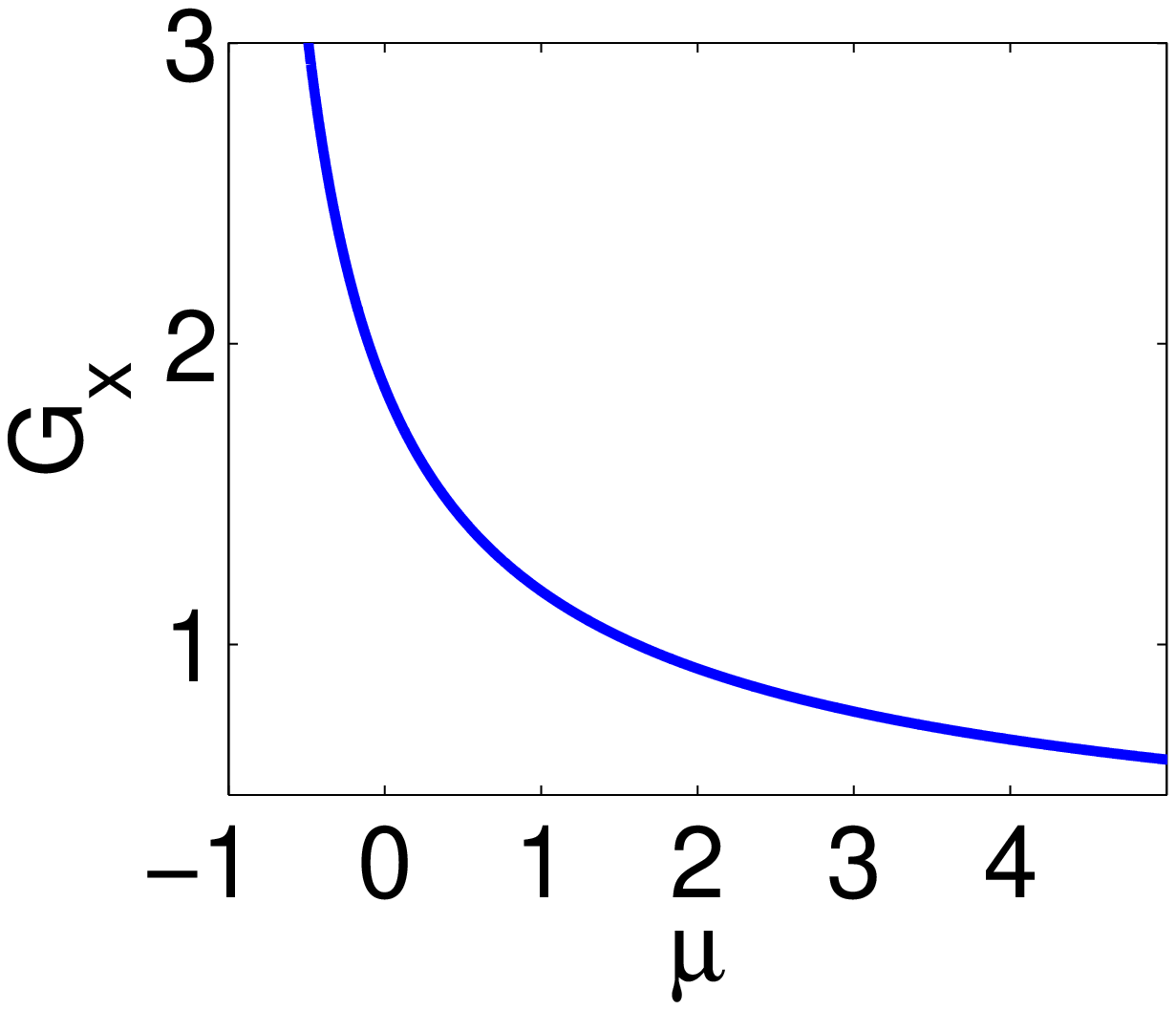}
\includegraphics[width=0.23\textwidth]{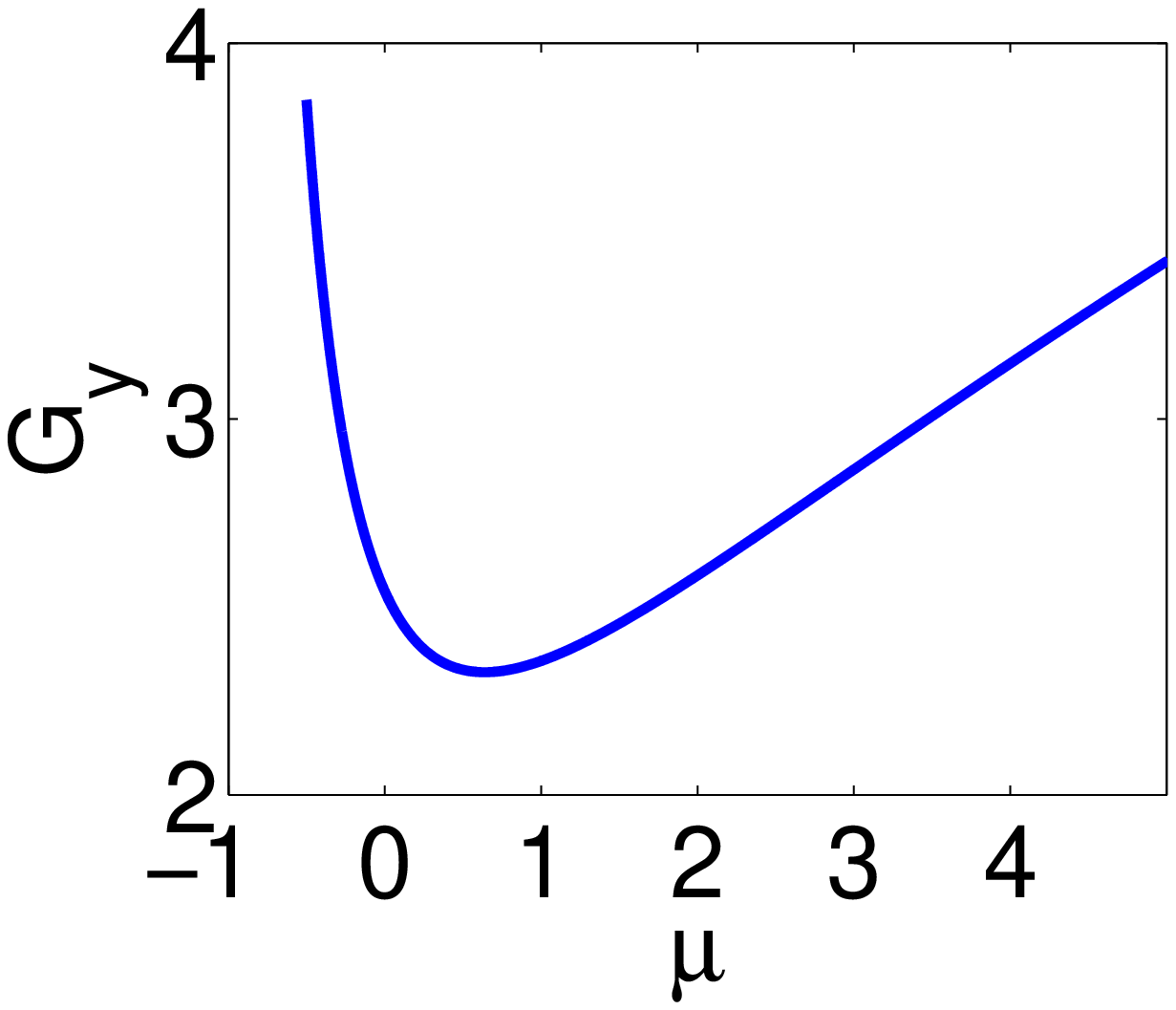}
\caption{(Color online) The functions $G_x\left(\mu\right)$ and $G_y\left(\mu\right)$ defined by Eq.~(\ref{Gx_def})-(\ref{Gy_def}). }\label{Fig_Functions}
\end{figure}

The functions $G_x(\mu)$ and $G_y(\mu)$ are shown in Fig.~\ref{Fig_Functions}. Fixing $\epsilon_t$ (see blue solid curve in Fig.~\ref{Fig_H0_T_Diagram}, the arrows correspond to the increasing of $\mu$) in order to exclude the trivial dependence of the fluctuation conductivity on $H_0$ due to the shift of the critical temperature one can see that the above functions correspond to the resulting dependencies $\left<\Delta\sigma^{\alpha\alpha}\right>(H_0)$. These dependencies differ qualitatively: $\left<\Delta\sigma^{yy}\right>(H_0)$ reveals a minimum at $H_0\sim H_{cr}$ due to strong damping of the velocity projection near $H_{cr}$ while $\left<\Delta\sigma^{xx}\right>(H_0)$ is a monotonically increasing function.

Note that the expressions (\ref{Inter_X_result}) and (\ref{Inter_Y_result}) are valid only in the vicinity of the critical field, i.e. for $H_0\approx H_{cr}$. If $H_0$ strongly differs from $H_{cr}$ then for a fixed shift from the critical temperature the fluctuation conductivity saturates as a function of $H_0$ due to the saturation of the effective mass component $m_y^{*-1}$. Indeed, for $H_0\ll H_{cr}$ the effective mass is given by the expression $m_y^{*-1}\approx m^{-1}$ (see Section \ref{Sec_Weak_Field}) and for $\epsilon_H=const$ the fluctuation conductivity is constant. In the opposite case for $H_0\sim H_{c2}^0$ the minimum of the energy spectrum shifts towards $k_y=\pm k_0$ and the effective mass is equal to $m_y^{*-1}\approx 0.58 m^{-1}$ (see Section \ref{Sec_Strong_Field}). As a result, for a fixed $\epsilon_{DW}$ the Aslamazov-Larkin correction to the conductivity does not depend on $H_0$ far from $H_{cr}$.

\section{Conclusion}\label{Sec_Conclusion}

In conclusion, we have shown that the fluctuation Aslamazov-Larkin conductivity tensor $\Delta\sigma$ of a hybrid structure consisting of a superconducting film and a ferromagnet with magnetic domains is essentially anisotropic. The magnitude of this anisotropy strongly depends on temperature as well as on the ratio between the width of domains $d$ and the magnetic length $L_{H_0}=\sqrt{\Phi_0/H_0}$, where $H_0$ is the amplitude value of stray magnetic field.

For $L_{H_0}\gg d$ the temperature dependence of $\Delta\sigma$ has a standard two-dimensional form $\Delta\sigma\propto\left(T-T_c\right)^{-1}$. At the same time, the conductivity tensor becomes slightly anisotropic: the ratio between the conductivity components along the domain walls and across them has the form
\begin{equation}
\frac{\sigma^{yy}}{\sigma^{xx}}=
1-\frac{2\pi^2}{15}\left(\frac{d}{L_{H_0}}\right)^4 \frac{\Delta\sigma}{\sigma_N},
\end{equation}
where $\sigma_N$ is the Drude conductivity due to the normal electrons and $\Delta\sigma$ is the Aslamazov-Larkin correction to the conductivity of the isolated superconducting film.

For $L_{H_0}\ll d$ the temperature dependencies of $\Delta\sigma$ components along and across the domain walls are essentially different. For $\epsilon_{DW}\gg \xi_0^2\left(dL_{H_0}^2\right)^{-2/3}$ the fluctuation conductivity tensor is isotropic. In the temperature range $\xi_0^2/d^2\ll \epsilon_{DW} \ll \xi_0^2\left(dL_{H_0}^2\right)^{-2/3}$ the component of the fluctuation conductivity across the domain walls is not singular at the critical temperature of domain-wall superconductivity while the component along domain walls reveals a crossover to a one-dimensional $\epsilon_{DW}^{-3/2}$ behavior. As a result, in this temperature range $\Delta\sigma$ becomes anisotropic and the ratio between the components along and across the domain walls is given by the relation  $\left(\xi_0^3/L_{H_0}^2d\right)\epsilon_{DW}^{-3/2}$. Finally, for $\epsilon_{DW}\ll\xi_0^2/d^2$ the $\Delta\sigma$ component across the domain walls reveals a crossover to the divergence $\Delta\sigma\propto\epsilon_{DW}^{-1/2}$. The resulting ratio between the components along and across domain walls does not depend on $d$ and is proportional to  $\left(\xi_0/L_{H_0}\right)^2\epsilon_{DW}^{-1}$.

In the intermediate case when $L_{H_0}\approx d$ the fluctuation conductivity has a peculiar dependence on the stray magnetic filed amplitude $H_0$: for fixed temperature shift from the transition temperature the dependence of the component of the conductivity tensor along the domain walls has a minimum at $H_0\sim \Phi_0/d^2$ while the transverse conductivity component is a monotonically increasing function of $H_0$.

We hope that the observation of above non-trivial fluctuation conductivity behavior in precise transport measurements above the critical temperature will provide more detailed information about peculiarities of the superconducting transition in superconductor/ferromagnet systems.

\section*{ACKNOWLEDGEMENTS}\label{Acknow}

The authors thank N.B. Kopnin, S.V. Sharov, I.A. Shereshevsky and A.Yu. Aladyshkin for useful discussions and suggestions. This work is supported by the
Russian Foundation for Basic Research,
RAS under the Federal Scientific Program ``Quantum physics of
condensed matter", Presidential RSS Council, the ``Dynasty'' Foundation, the Russian Agency
of Education under the Federal Program ``Scientific and
educational personnel of innovative Russia in 2009--2013".

\section*{APPENDIX}\label{Sec_Appendix}

To calculate the fluctuation conductivity in case $H_0\approx\Phi_0/d^2$ let us substitute the spectrum (\ref{Inter_Spectrum}) into Eq.~(\ref{startexpr}). The diagonal matrix elements of the velocity operator have the form
\begin{equation}\label{Inter_velocity}
v^x=2\alpha k_x,~~~~v^y=2\beta k_y +4\gamma k_y^3,
\end{equation}
where $\alpha=\left(m/m_x^{*}\right)\xi_0^2$, $\beta=\left(m/m_y^{*}\right)\xi_0^2$, $\gamma=4m\xi_0^2\eta$. Then the expression for $\left<\Delta\sigma^{xx}\right>$ takes the form
\begin{equation}\label{App_X_start}
\left<\Delta\sigma^{xx}\right>=\frac{e^2}{4\pi\hbar w_s}\int\limits_0^{\infty}dk_y\int\limits_0^{\infty}dk_x \frac{4\alpha^2 k_x^2}{\left(\epsilon_t+\alpha k_x^2+\beta k_y^2+ \gamma k_y^4\right)^3},
\end{equation}
Integrating this expression over $k_x$ we obtain:
\begin{equation}\label{App_X_int_Kx}
\left<\Delta\sigma^{xx}\right>=\frac{e^2\sqrt{\alpha}}{16\hbar w_s}\int\limits_0^{\infty} \frac{dk_y}{\left(\epsilon_t+\beta k_y^2+ \gamma k_y^4\right)^{3/2}}=-\frac{e^2\sqrt{\alpha}}{8\hbar w_s}\frac{\partial I_0}{\partial \epsilon_t},
\end{equation}
where
\begin{equation}\label{App_UniInt}
I_0=\int\limits_0^{\infty} \frac{dk_y}{\sqrt{\epsilon_t+\beta k_y^2+ \gamma k_y^4}}.
\end{equation}
It is convinient to introduce the parameter $\mu=\beta/2\sqrt{\gamma\epsilon_t}$, which for $T>T_c$ takes the values from the interval $-1<\mu<\infty$. Then there are two different cases. The first one is $-1<\mu<1$. In this case the integral $I_0$ can be rewritten in the form
\begin{equation}\label{App_UniInt_1}
I_0=\frac{1}{\sqrt{\gamma}}\int\limits_0^{\infty} \frac{dk_y}{\sqrt{\left(k_y^2+\rho^2\right) \left(k_y^2+\rho^{*2}\right)}}.
\end{equation}
Here $\rho^2=\sqrt{\epsilon_t/\gamma} \left(\mu+i\sqrt{1-\mu^2}\right)$ and the asterisk indicates complex conjugation. Then the result is \cite{Prudnikov}
\begin{equation}\label{App_UniInt_Res_1}
I_0=\frac{1}{\sqrt{\gamma}\left|\rho\right|}K\left(\frac{{\rm Im}^2\left(\rho\right)}{\left|\rho\right|^2}\right),
\end{equation}
where $K\left(\nu\right)$ is the complete elliptic integral of the second kind.

In case $\mu>1$ the integral can be represented in the form
\begin{equation}\label{App_UniInt_2}
I_0=\frac{1}{\sqrt{\gamma}}\int\limits_0^{\infty} \frac{dk_y}{\sqrt{\left(k_y^2+a^2\right) \left(k_y^2+b^2\right)}},
\end{equation}
where $a^2=\sqrt{\epsilon_t/\gamma}\left(\mu+\sqrt{\mu^2-1}\right)$, $b^2=\sqrt{\epsilon_t/\gamma}\left(\mu-\sqrt{\mu^2-1}\right)$. Then the result is \cite{Prudnikov}
\begin{equation}\label{App_UniInt_Res_2}
I_0=\frac{1}{\sqrt{\gamma}a} K\left(\frac{\sqrt{a^2-b^2}}{a}\right).
\end{equation}

After substitution of the expressions (\ref{App_UniInt_Res_1}) and (\ref{App_UniInt_Res_2}) into Eq.~(\ref{App_X_int_Kx}) and taking derivative over $\epsilon_t$ we obtain the expression (\ref{Inter_X_result}).

Analogously the component $\left<\Delta\sigma^{yy}\right>$ of the fluctuation conductivity tensor can be written as
\begin{equation}\label{App_Y_start}
\left<\Delta\sigma^{yy}\right>=\frac{e^2}{4\pi\hbar w_s}\int\limits_0^{\infty}dk_y\int\limits_0^{\infty}dk_x \frac{\left(2\beta k_y+4\gamma k_y^3\right)^2}{\left(\epsilon_t+\alpha k_x^2+\beta k_y^2+ \gamma k_y^4\right)^3}
\end{equation}
Integrating this expression over $k_x$ we obtain
\begin{equation}\label{App_T_int_Kx}
\begin{array}{c}{\ds
\left<\Delta\sigma^{yy}\right>=\frac{e^2}{16\sqrt{\alpha}\hbar w_s} \int\limits_0^{\infty} \frac{\left(\beta+6\gamma k_y^2\right)dk_y}{\left(\epsilon_t+\beta k_y^2+ \gamma k_y^4\right)^{3/2}}=}\\{\ds ~~~~~~~= -\frac{e^2\beta}{8\hbar\sqrt{\alpha} w_s}\frac{\partial I_0}{\partial \epsilon_t}-\frac{3e^2\gamma}{4\hbar\sqrt{\alpha} w_s}\frac{\partial I_0}{\partial \beta},}
\end{array}
\end{equation}
where $I_0$ is defined by Eq.~(\ref{App_UniInt}). Then using the expressions (\ref{App_UniInt_Res_1}) and (\ref{App_UniInt_Res_2}) one can obtain the result (\ref{Inter_Y_result}).

\end{document}